\documentclass[journal=jacsat,manuscript=article]{achemso}
\usepackage{chemformula} 
\usepackage[T1]{fontenc} 

\author{Sutapa Ghosh}
\email{sutapa.g@campus.technion.ac.il}
\affiliation[Unknown University]
{Andrew and Erna Viterby Department of Electrical Engineering and Russell Berrie Nanotechnology Institute, Technion-Israel Institute of Technology, Haifa 32000, Israel}
\author{Gadi Eisenstein}
\title{A phase stable hybrid dual comb spectrometer}

\abbreviations{DCS, FC, MLL}
\keywords{Dual comb spectroscopy, frequency comb, mode-locked lasers}

\begin{document}

\begin{tocentry}
\includegraphics[scale=0.25]{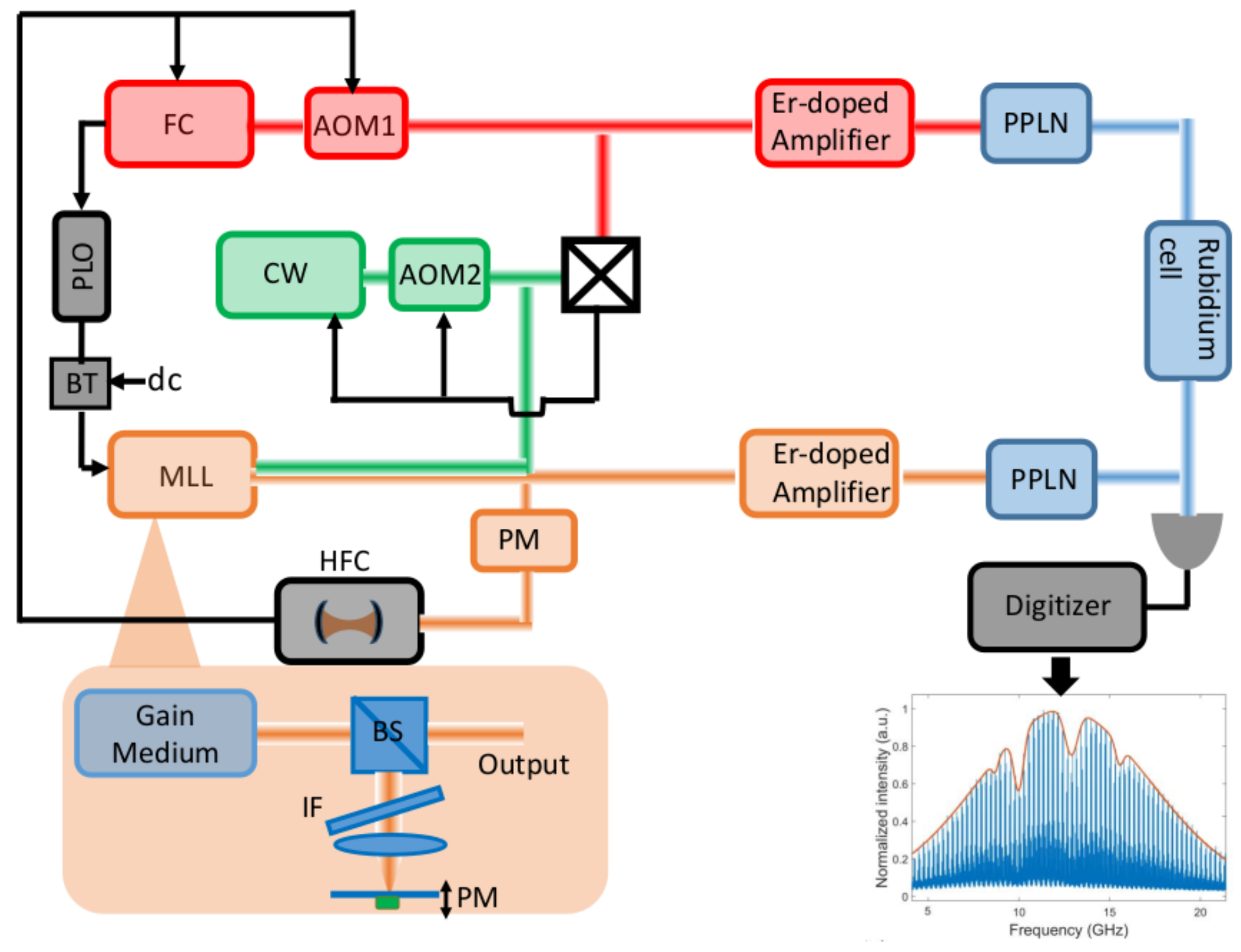}\\

Manuscript title: "A phase stable hybrid dual comb spectrometer"\\

Authors: Sutapa Ghosh and Gadi Eisenstein\\

Brief synopsis:  A hybrid dual comb spectrometer with high absolute stability and long mutual coherence time is presented.\\

\end{tocentry}

\begin{abstract}
Dual comb spectroscopy (DCS) is a broadband technique offering high resolution and fast data acquisition. We describe a hybrid dual comb spectrometer comprising a broadband commercial fiber laser system offering a wide range of sample interrogation, and an actively mode locked semiconductor laser (MLL) having a widely tunable, relatively narrow spectrum. The mutual coherence over 100 seconds has been realized between the two combs. We employed the DCS system to characterize the absorption spectrum of rubidium atoms at 313 K with a high signal to noise ratio. The broadband laser is directly locked on a high finesse cavity, providing long-term stability, while the semiconductor laser is locked to it.  To characterize the absolute stability of the DCS system, the linewidth of the MLL comb line is measured and shown to reduce from 880 kHz to $17$ kHz when the system is fully locked. The long-term stability was measured to be $5 \times 10^{-12}$ at $1$ second and $5 \times 10^{-14}$ at $350$ seconds. The measured timing jitter of the MLL is ten times smaller due to the overall locking. In addition, we have addressed the effect of dispersion on the locking quality, which is significant for broadband comb lasers.
\end{abstract}

\section{Introduction}
The invention of broad band coherent frequency combs (FC) \cite{udem_99,diddams_01,udem_02} has enabled new schemes for broad band spectroscopy \cite{diddams_07,picque_19} at a wide range of wavelengths. The availability of highly stable FC has enabled a related, highly versatile spectroscopy scheme known as dual comb spectroscopy (DCS) \cite{keilmann_04,schliesser_05,coddington_16}. DCS offers significant advantages over other broadband spectroscopy methods mainly in terms of high frequency resolution, fast data acquisition and mechanical stability due to the absence of any moving part \cite{newbury_10}. DCS involves two frequency combs with slightly different repetition rates. Generally, one laser interrogates a sample such that the properties of the sample get mapped onto that laser spectrum, which is consequently sampled by the second laser, generating beats in the RF domain at multiples of the difference between the repetition rates. The sample properties are detectable on the RF spectrum. A crucial property of a DCS system is the mutual coherence of the two combs which must be long enough to allow data acquisition to characterize the weak molecular interactions. Systems with various degrees of mutual coherence ranging from two free running \cite{burghoff_16,dutt_18,hebert_17,sterczewski_19} to phase locked combs (with and without locking to an external reference)\cite{coddington_08,villares_14}, having different spectral resolutions, have been demonstrated. 

The most common FC laser sources are fiber-based or Ti:Sa mode locked lasers with an octave-wide spectrum that enables the well-known f-2f stabilization scheme of the carrier envelop offset \cite{telle_99,helbing_02,d_jones_00}. The repetition rate is stabilized by locking the laser cavity length either to an optical reference or to a stable radio-frequency (RF) source \cite{jones_01}. To attain mutual coherences between two combs, a stable CW laser can be used to synchronize both lasers separately. With this approach, mutual coherences up to 1s can be attained. Further enhancement in the mutual coherence time have been achieved using various numerical means \cite{zolot_12,burghoff_16}. With feed forward relative CEO stabilizations, mutual coherences of up to 2000 seconds have been reported \cite{chen_18}.

An interesting DCS structure where a single semiconductor mode-locked integrated external-cavity surface emitting laser (MIXSEL) that generates two combs at slightly different repetition rates was demonstrated by Keller and co-workers~\cite{link_17}. These two combs are naturally highly coherent what avoids all locking electronics. Nevertheless, the laser cavity and the pump laser need to be stabilized with respect to an absolute reference such as an optical cavity or some atomic transition to ensure long term stability. The MIXSEL scheme emits relatively narrow spectra which results in high powers per spectral line leading to high signal to noise ratios but is limited in the chemical species it can interrogate. There are also efforts to develop hybrid DCS systems that combine different laser sources. One such DCS configuration was realized with a quantum cascade laser and a fiber-based mode locked lasers to take advantages of both platforms \cite{consolino_20}. However, it is challenging to establish high mutual coherences between such vastly different lasers.

Different DCS schemes have been used to perform numerous spectroscopic experiments studying many materials~\cite{kippenberg_11,long_14,mateos_15,duran_15,millot_16,yin_14} thereby establishing the technique as a major tool for broadband spectroscopy. 

This paper describes a new hybrid DCS system that exhibits a very long, up to 100 seconds, mutual coherence time between the two lasers. It combines the advantage of both a broadband fiber based laser and a narrowband active mode-locked semiconductor lasers (MLL). The sample is interrogated with the broadband laser which allows to perform DCS over a wide range of samples while the spectrum is extracted by mixing the a FC with the widely tunable MLL whose power in each line is high. Since the MLL is locked on a single line of the FC, it can be tuned over the FC spectrum and re-stabilized on FC. With combination of various locking schemes, the DCS system yield a high long-term absolute stability and the long mutual coherence.

The system we devised was used to measure the absorption spectrum of rubidium atoms and showed a mutual coherence time of the system up to 100 seconds without any numerical phase correction. Moreover, we have stabilized the absolute frequency of the DCS setup by measuring the frequency fluctuation of the MLL in the transmission mode of a high finesse cavity, while it is locked to the FC, and used the measurement to correct the frequency of the FC. This resulted in a large reduction of the linewidth of the MLL from 880 kHz to 17 kHz when the system was fully locked. The long-term stability of the system was evaluated in terms of the Allan deviation and was shown to be $5 \times 10^{-12}$ at 1 second and $5 \times 10^{-14}$ at 350 seconds. Time domain characterization of the MLL (whose repetition rate was 250 MHz) yielded a jitter of 0.13 ps. The high absolute, long-term stability and the long mutual coherence of the system makes it suitable for Doppler-free spectroscopy of a wide range of weak transition materials.

\begin{figure}[h]
\begin{center}
\includegraphics[scale=0.5]{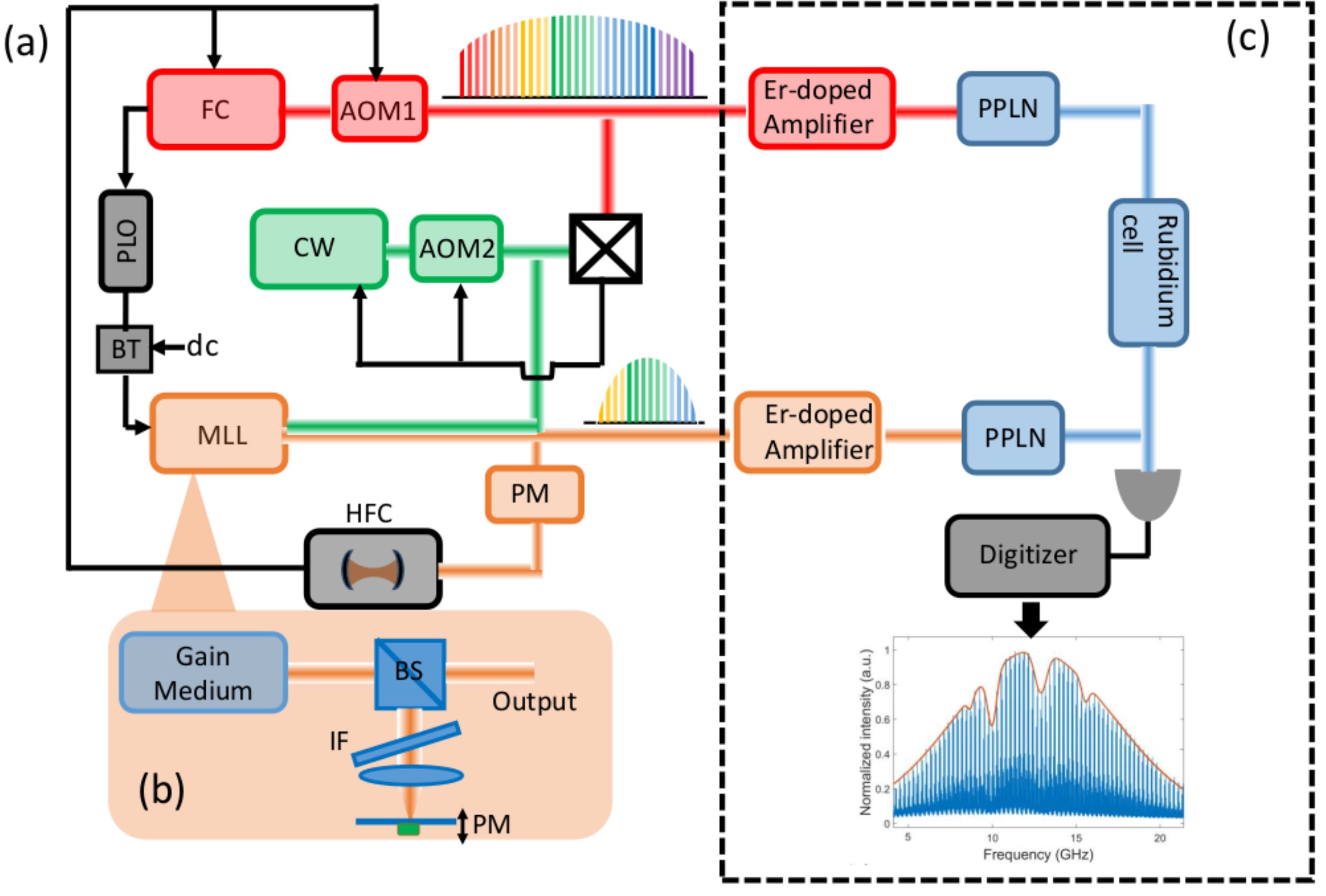}
\caption{{\bf Schematics of the hybrid set-up for dual-comb spectroscopy.} (a) A commercial FC is used to stabilize a CW  laser which injection locks the MLL. This transfer the fluctuations of the FC to the carrier envelop offset (CEO) of the MLL. The repetition rate of the MLL is determined by the modulation frequency from a phase-locked oscillator (PLO) referenced on an RF signal derived from the FC repetition rate. The FC is stabilized to a high finesse cavity which provides long term stability. AOM: Acoustic-optic modulator, BT: Bias-Tee, PM: Phase modulator, HFC: high finesse cavity (b) A detailed schematic of the MLL is shown in the orange shaded area. BS: Non-polarizing beam splitter, IF: Interference filter, PM: Piezo mounted mirror. (c) Both FC and MLL are amplified and frequency doubled by a periodically poled lithium niobate crystal (PPLN). The frequency doubled FC interrogates the rubidium cell and is combined with the second harmonic of the MLL to generate the RF beats.  }
\label{fig1}
\end{center}
\end{figure}

\section{Hybrid dual comb spectrometer}

Fig.1 describes an overview of the hybrid dual comb spectroscopy apparatus. A commercial fiber laser FC (Menlo systems, FC1500-250-ULN) with a repetition rate, $f_{rep}$ of 250 MHz is one of the lasers. The second is an active mode locked laser, MLL whose repetition rate is detuned from the FC by $\delta f_{rep}$ of 25 kHz and is shown in the orange shaded part of Fig.1 (b). It employs a piezo-mounted mirror in the cat-eye configuration and an intracavity interference filter. The MLL is injection locked by a CW external cavity semiconductor laser which controls the carrier envelop offset (CEO) of the MLL. The CW laser is stabilized on a single tooth of the FC. The beat between the CW laser and the FC is used to stabilize the CW laser by applying a fast-feedback to an acousto-optic modulator, AOM2 and a slow feedback to the piezo mounted cavity mirror. This allows the CEO of the MLL to follow the frequency fluctuation of the FC what eliminates the need to stabilized the CEO using the conventional f-2f scheme \cite{telle_99,helbing_02,d_jones_00} which requires  an octave wide spectrum. The repetition rate of the MLL is set by its RF drive drive whose frequency is derived from the FC repetition rate. This allows the MLL to continuously follow the timing and phase fluctuations of the FC in real-time, yielding the long coherence times between the two combs. 

The MLL was used to generate the error signal since its power per line is larger than that of the FC (for the same average power). The MLL was phase modulated at 12 kHz by an electro-optic modulator and transmitted through the cavity at the output of which it was demodulated by a lock-in amplifier. The error signal feeds the feedback control of the FC CEO through AOM1 and provided slow feedback to the length of the FC. The fully stabilized MLL emits pulses with a duration of 20 ps. 

The dual-comb interferometer was used to measure the Doppler broadened spectrum of the rubidium atoms at 313 K. Both the FC and MLL were tuned to 1560 nm and were amplified by an erbium doped amplifier. The amplified signals were frequency doubled by a temperature stabilized periodically poled lithium niobate crystals to generate a second harmonic at 780 nm. The frequency doubled FC was focussed through the 75 cm glass cell with rubidium atoms held at 313 K. After interacting with the gas cell in a single-pass configuration, the frequency-doubled FC was combined with the frequency-doubled MLL and detected by a photo-detector, PD, FPD610-FC-VIS. The detected signal was filtered, amplified and digitized. A fast Fourier transform was computed and the amplitude of the spectrum was retrieved. A detailed outline of the experimental set-up is presented in the supplementary section I (Fig. S1).

\section{Results}
\subsection{Stabilization and characterization of the hybrid DCS}

The phase-modulated MLL is transmitted through the cavity and demodulated by a lock-in amplifier. Around the cavity resonance, the phase of the laser lines changes by $\pi$ yielding a phase difference between the resonant mode and its modulated components forming a dispersive error signal with a zero-crossing as shown in Fig. 2 (a). This error signal fed the feedback control of the FC CEO through AOM1 and provided slow feedback to the length of the FC. The free spectral range of the cavity is 6 GHz, which means that every 24th mode of the MLL passes through the cavity and contributes to the error signal. Fig. 2(b) shows the slope of the error signal as a function of the modulation frequency and modulation depth. The values used in the experiment are represented by a red star in the figure. Stabilization using the transmission mode was previously demonstrated for CW lasers~\cite{hils_87} while for a pulsed laser, a single line was used to lock the repetition rate of the entire comb~\cite{kruger_95}. The present system has contributions of multiple comb lines stabilizing the entire comb spectrum with a high signal to noise ration (SNR).

\begin{figure*}[!h]
\includegraphics[scale=0.4]{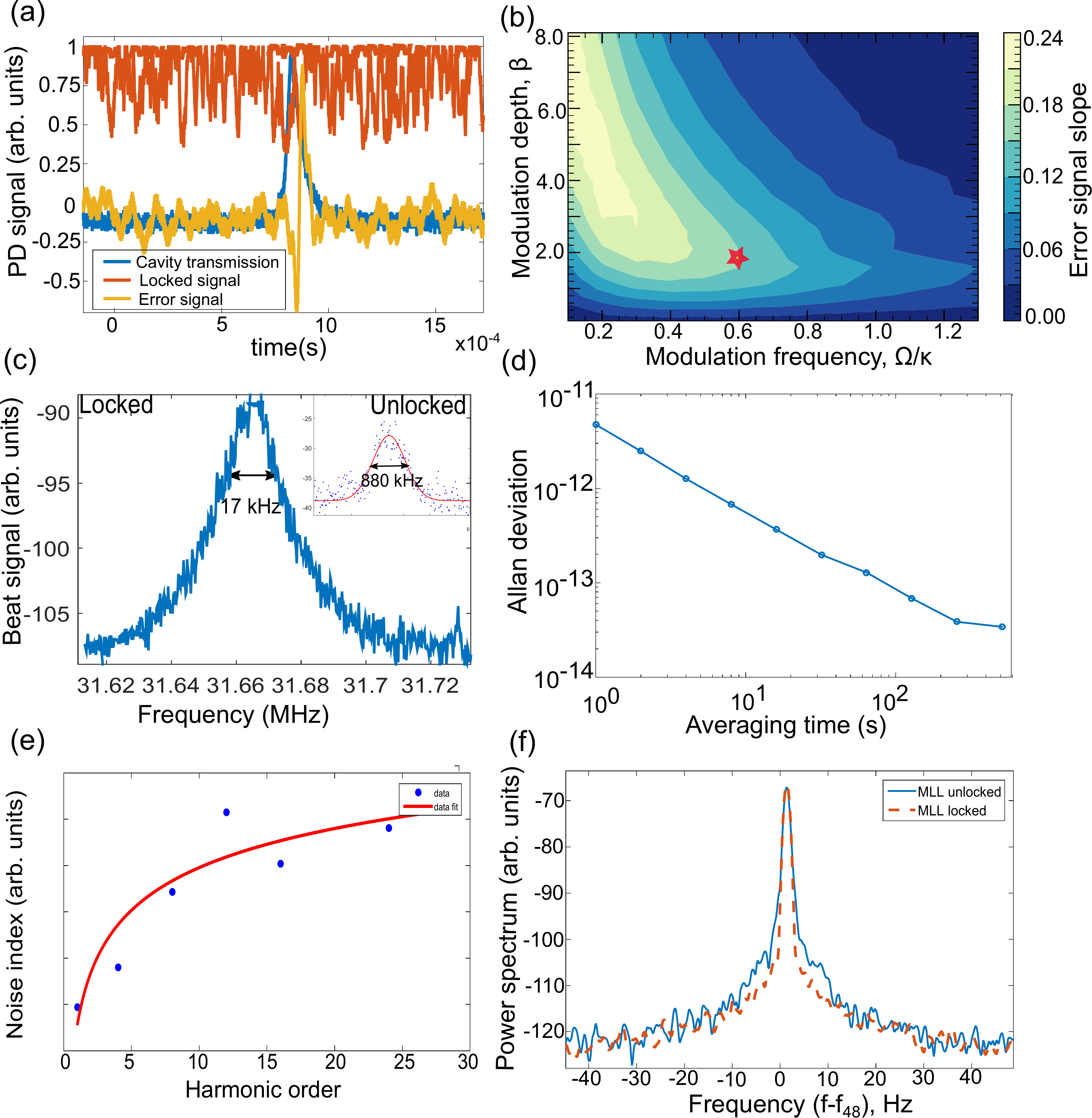}
\caption{{\bf Error signal stabilizing the MLL.} (a) Measured  cavity transmission  (blue), error signal (yellow) and locked cavity transmission signal (red). (b) The slope of the error signal as a function of modulation frequency $\Omega$ and modulation depth $\beta$. A red star signifies the experimentally used parameters. (c) The heterodyne beat between the MLL and a cavity stabilized CW laser characterizing the MLL stability. The linewidth of the MLL which represents the short-term frequency noise reduced from 880 kHz in the free running case (shown in the inset) to 17 kHz for the fully locked system. (d) The Allan deviation showing long term stability of $5 \times 10^{-12}$ at 1 second and $5 \times 10^{-14}$ at 350 seconds. (e) Timing jitter of the MLL, measured in terms of the dependence on harmonic number of the noise index which is the ratio of the integrated noise skirt to the peak power of the corresponding harmonic.  (f) The power spectrum of the 48th harmonic under locked and unlocked conditions.}
\label{fig2}
\end{figure*}

To quantify the quality of our locking scheme, we performed a heterodyne measurement of the MLL with a second CW laser which was locked to the high finesse cavity and has a linewidth below 100 Hz. This yielded the short-term frequency noise. For the free-running MLL, the linewidth of the heterodyne beat was around 880 kHz while under a stabilized condition, the linewidth reduced to 17 kHz as shown in Fig. 2 (c). The long-term stability was characterized in terms of the Allan deviation (AD)~\cite{allan_66}. The heterodyne beat frequency between the MLL and the CW laser was recorded over 500 seconds to calculate the AD as shown in Fig. 2(d). The stability was found to be $5 \times 10^{-12}$ at 1 second and $ 5 \times 10^{-14}$ at 350 seconds.

The stability of the MLL was also characterized in the time domain by measuring its timing jitter using the Van der Linde technique \cite{linde_86}. The timing jitter of a pulse train exhibits a noise skirt around the peak of each harmonic of the detected signal. The noise index for each harmonic is defined as the ratio between the integrated power of the noise bands and the peak power of the corresponding harmonic.  Fig. 2 (e) shows the noise index as a function of the harmonic number. As the amplitude noise contribution grows linearly with the harmonic number $n$ while the frequency noise grows as $n^{2}$, the measurement accuracy increases with the harmonic number. Fig. 2(f) shows the power spectrum of the 48th harmonic (recorded with a resolution bandwidth of 1 Hz) of an unlocked and a locked MLL. The calculated timing jitter of the locked MLL (whose repetition rate was 250 MHz) was 0.13 ps which is more than a ten-fold improvement over the unlocked case.

\subsection{Effect of the cavity dispersion on stabilization}
The spectral position of the resonances in any laser shifts from its equidistant position due to the cavity dispersion \cite{jones_01}. This shift results in the broadening of the error signal what reduces its slope, when the contributions from all the modes are combined. Stabilization in transmission mode is more sensitive to the cavity dispersion as the carrier and modulated signal are spectrally very close to each other.

To study this effect, we consider a frequency comb field with a repetition rate, $\omega_{rep}$ and carrier-envelope offset, $\omega_{ceo}$. The electric field can be written as,
 
\begin{equation}
E_{in} = \sum_{l=-\infty}^\infty A_l e^{-i\omega_l t} + A_l^* e^{i\omega_l t}
\end{equation} 
 where, $\omega_l = l \omega_{rep} + \omega_{ceo}$, $\omega_l$ is the carrier frequency. For a comb modulated at frequency $\Omega$ with a modulation depth, $\beta$, the error signal is calculated as:
 
\begin{equation}
  \begin{aligned}
      P_{error} &= \sum_l A_l [\text{dc term} + J_0(\beta) J_1(\beta) \text{Re}\{T_l(\omega_l)T_l^*(\omega_l-\Omega) - T_l(\omega_l+n\Omega)T^*(\omega_l)\} + \\
      &\text{Im}\{T_l(\omega_l)T_l^*(\omega_l-\Omega) -     T_l(\omega_l+n\Omega)T^*(\omega_l)\} + \\
      &\sum_{n\neq m}\sum_{m=n-1}^{n+1} J_n(\beta)J_m(\beta) \text{Re}\{T_l(\omega_l-n\Omega)T_l^*(\omega_l-m\Omega) - T_l^*(\omega_l+n\Omega) T_l(\omega_l+n\Omega)\} + \\
          &\text{Im}\{T_l(\omega_l-n\Omega)T_l^*(\omega_l-m\Omega) - T_l^*(\omega_l+n\Omega) T_l(\omega_l+n\Omega)\}]
  \end{aligned}
\end{equation} 

where, $J_n(\beta)$ is the Bessel function of order $n$ and $T_l(\omega)$ is the transmission function corresponding to the $l^{th}$ comb line. Details of the calculation are presented in the supplementary section II. The slope of the error signal as a function of modulation frequency and depth, is plotted in Fig. 2(b) . To study the impact of dispersion on the error signal, we have added a mode-dependent shift, $\Delta_l$ to the cavity resonance frequency $\omega_{cav,l}$. The cavity transmission function $T_l(\omega)$ for the $l^{th}$ comb line is written as,

\begin{equation}
\begin{aligned}
T_l(\omega) = \frac{\kappa_{cav}[\kappa_{cav} - i(\omega - \omega_{cav,l}-\Delta_l)]}{(\omega - \omega_{cav,l}-\Delta_l)^2 + \kappa_{cav}^2}
\end{aligned}
\end{equation}

\begin{figure*}[t]
\includegraphics[width=\textwidth]{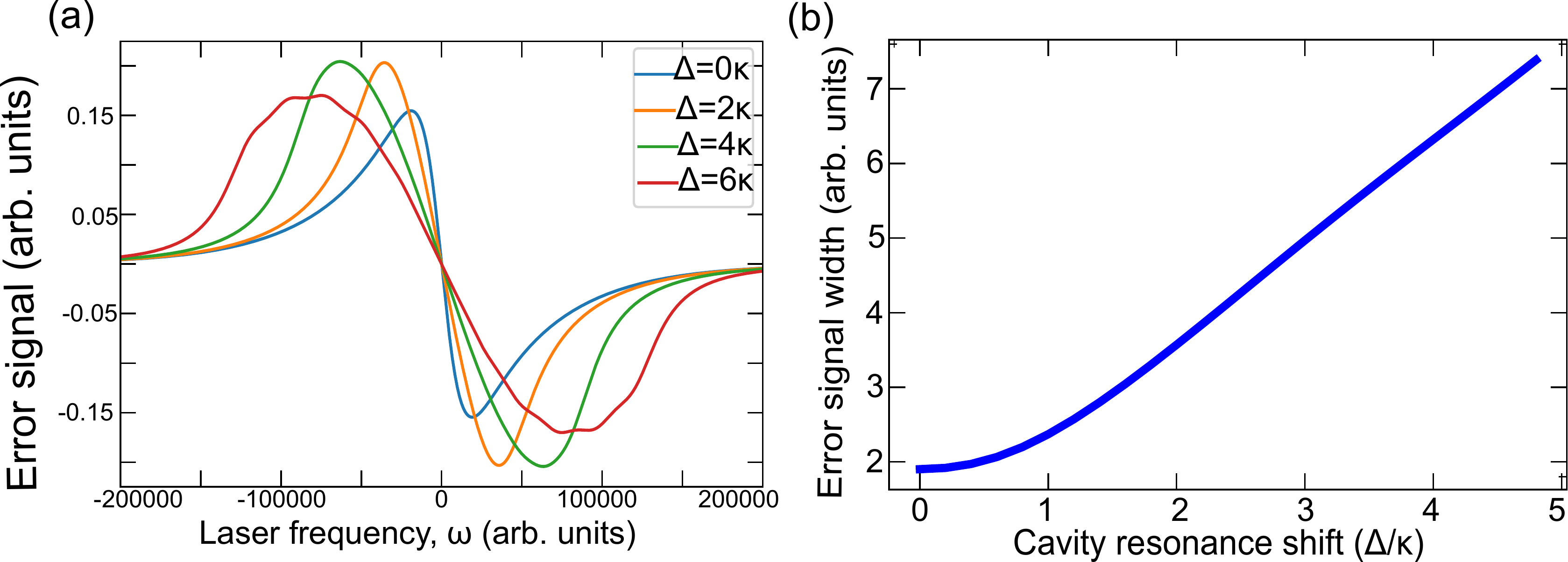}
\caption{{\bf Cavity dispersion effect on the error signal in a broadband comb laser.} (a) The cavity dispersion induces a linear shift of the cavity mode spacing from its equidistant position. The figure describes the calculated error signal as a function of the cavity mode shift, $\Delta$ induced by dispersion. $\kappa$ represents the 20 kHz cavity linewidth. (b) Calculated error signal width as function of the cavity resonance shift ($\Delta$).}
\label{fig3}
\end{figure*}

Depending on the level of dispersion, a linear shift of the resonance position as a function of line number, from the central line is added in the calculation. The error signal for different values of 
frequency shift ($\Delta_l$) resulting from the dispersion, is shown in Fig. 3 (a); the slope of the error signal decreases with the level of dispersion. The dependence of error signal width on the level of dispersion is presented in Fig. 3 (b).

The present experiment used the relatively narrow band, 25 GHz, MLL to generate the error signal and hence the role of dispersion was negligible.

\subsection{Dual comb spectroscopy with the rubidium atoms}

The mutual coherence of the system we devised was characterized by measuring the rubidium Doppler spectrum at 313 K with different integration times. The frequency doubled FC was focused on the rubidium cell and combined with the frequency doubled MLL.  The beat between them reveals the Doppler profile of the rubidium atoms as shown in Fig. 4(a). Various transition peaks of rubidium isotopes are isolated and shown in Fig.4 (b). Due to the technical limitations of the data acquisition system, the averaging was performed on the Fourier domain data. A Doppler width of 535 MHz (FWHM) is extracted from the absorption spectra of $^{87}\text{Rb}$ and $^{85}\text{Rb}$ after averaging for 100 seconds, as described in Fig. 4(c). The calculated SNR for different integration times is shown in Fig. 4 (d). Mutual coherence leads to the slope of 0.51 confirms the square root dependence of the SNR on the integration time \cite{coddington_16}.

\begin{figure*}[t]
\begin{center}
\includegraphics[width=\textwidth]{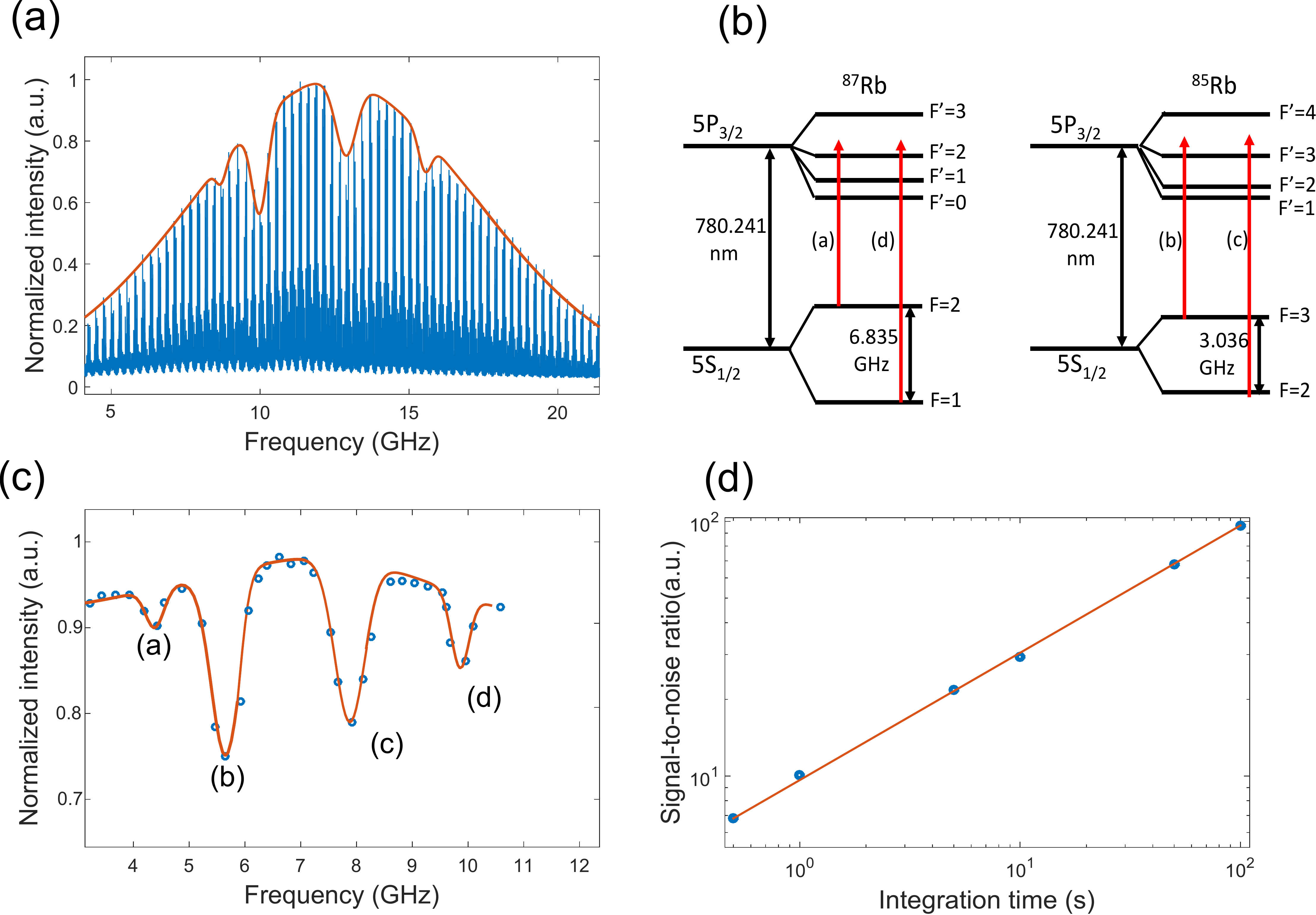}
\caption{{\bf Dual-comb spectroscopy of rubidium atom.} (a) The recorded RF beats for an averaged data integration time of 100s. The Doppler broadened spectrum of rubidium atom is extracted from the beats. b) Energy level diagram of $^{87}\text{Rb}$ and $^{85}\text{Rb}$. (c) Rubidium spectrum with an extracted Doppler width of 535 MHz (FWHM) at 313 K. The spectrum is obtained from the beat profile after subtracting the background originating from the narrowband comb laser. (d) as a function of different data integration times. The slope of 0.51 indicated the expected square root behavior of the SNR on the measurement time.}
\label{fig4}
\end{center}
\end{figure*}

Mutual coherence is independent of the absolute stability of the FC. However, long data acquisition times demand that the long-term FC frequency fluctuations should be much lower than the sample absorption width so as not to degrade the SNR. This limitation is severe for Doppler free spectroscopy. The direct locking of a pulsed laser to a high finesse cavity, which we have implemented, simplifies the stabilization set up and ensures  high, long term absolute stability. Since the error signal is generated in the transmission mode, the restriction that all comb laser lines have to be resonant with the cavity mode (which is the case in reflection mode) is eliminated. The repetition rate of the comb is tuned such that some lines at the spectrum edges are resonant with the cavity mode and contribute to the error signal. This lead to the stabilization of entire comb spectrum.

\section{Discussion}

In conclusion, we have constructed a hybrid dual comb spectrometer with a broadband fiber comb laser and an injection locked, MLL. Different locking schemes established mutual coherence of 100 seconds between them. The maximum measured mutual coherence time is set by limitations of the electronic digitizer whose storage capacity is insufficient for measurements longer than 100 second. We have demonstrated a method to reference the DCS set up to a high finesse cavity reaching a stability of $5 \times 10^{-12}$ at 1 second and $5 \times 10^{-14}$ at 350 seconds. The stability can be increased by improving the bandwidth of the cavity lock. Moreover, we have addressed the impact of cavity dispersion on the stabilization of broad band sources. We have shown that the dispersion broadens the error signal when the contribution of far comb lines are combined, reducing the stabilization quality. However, generating several error signals from different comb spectral regions can mitigate the effect of cavity dispersion as this limits the number of modes contributing to a single error signal and reduces the impact of the broadband spectrum.

The DCS set up can in principle be simplified by using two narrow band MLL lasers with a single CW laser to injection lock them. In this way, the lasers will be mutually phase coherent without the need for any active stabilization. However, the CW laser must be stabilized in order to attain absolute long term stability. This can be done, for example, by the conventional Pound-Drever-Hall locking scheme \cite{black_01}.

\section{Notes}
The authors declare no conflicts of interest.

\begin{acknowledgement}
This work was partially supported by the PMRI-Peter Munk Research Institute - Technion.
\end{acknowledgement}

\begin{suppinfo}
The supporting document (PDF) involves the detailed experimental set-up and the calculations for the results used in the main article.
\end{suppinfo}

\bibliography{MLL}

\end{document}